\documentclass[a4paper,10pt]{article}

\usepackage{amssymb,amsmath,color,cite,graphicx,caption,subcaption}

\title{Reduction scheme for coupled Dirac systems}
\author{Miguel Castillo-Celeita$^{1,2}$, V\'i{}t Jakubsk\'y$^{1}$\\
	$^1$\textit{The Czech Academy of Science, Nuclear Physics Institute, \v Re\v z/Prague, Czech
		Republic}\\
	$^2$\textit{Physics Department, Cinvestav, P.O. Box. 14-740, 07000 Mexico City, Mexico}\\ }

\begin{document}
	\maketitle
	
	\begin{abstract}
		We analyze a class of coupled quantum systems whose dynamics can be understood via two uncoupled, lower-dimensional quantum settings with auxiliary interactions. The general reduction scheme, based on algebraic properties of the potential term, is discussed in detail for two-dimensional Dirac Hamiltonian. We discuss its possible application in description of Dirac fermions in graphene or bilayer graphene in presence of distortion scattering or spin-orbit interaction. We illustrate the general results on the explicit examples where the involved interactions are non-uniform in space and time.
	\end{abstract}
\section{Introduction}


In this article, we focus on two coupled quantum systems that are  described by $2+1$-dimensional Dirac equation. We are particularly interested in the physical scenarios that can appear in physics of graphene and other Dirac materials.  
It is well known that the valence and conduction bands of graphene get in touch in the corners of the first Brillouin zone. The dispersion relation is linear in the vicinity of these points. Only two of them, $K$ and $K'$, are inequivalent and give rise to two generations of two-dimensional Dirac fermions \cite{Semenoff}. The fermions both from  $K$-valley and $K'$-valley enjoy pseudo-spin degree of freedom associated with presence of two triangular sublattices in the crystal. Additionally, they possess spin-degree of freedom. Therefore, the Hamiltonian of Dirac fermions in graphene should have the form of $8\times8$ matrix operator.  Nevertheless, only the degrees of freedom relevant for the considered interactions  are usually taken into account. 

For instance, disorder in the lattice can lead to the intervalley scattering. The Dirac fermions can get scattered from $K$-valley into $K'$-valley and vice versa by atomic defects, deposits on the crystal or by the mechanical contacts with the substrate. The general Hamiltonian $H_{dis}$ for disorder scattering  of Dirac fermions in graphene can be written in the following form \footnote{We set the Fermi velocity in graphene $v_F=1$ throughout the article.} \cite{Altland}, \cite{Kechedzhi}, \cite{Manes}, see also \cite{Shon}, \cite{AndoDisorderScatt}, \cite{Frassdorf}, \cite{AndoCross}, 
\begin{equation}\label{Hdis0}
H_{dis}=\left(\begin{array}{cccc}
V_A&\pi+V&W_A&W^{+}\\
\pi^\dagger+\overline{V}& V_B&W^{-}&W_B\\
\overline{W}_A&\overline{W}^{-}&V_A& -\pi^\dagger+V'\\
\overline{W}^{+}&\overline{W}_B&-\pi+\overline{V'}&V_B
\end{array}\right),\quad \pi=-i\partial_x-\partial_y. 
\end{equation}
The bispinor components are ordered here as $(A_K,B_K,A_{K'},B_{K'})$ where $A(B)$ denotes the pseudo-spin and $K(K')$ the valley degree of freedom \footnote{For instance, the wave function $\Psi=(\psi,0,\xi,0)$ would be located just on the triangular sublattice $A$.}. The interaction in (\ref{Hdis0}) preserves the spin of the Dirac fermions. 

Spin-orbit coupling does not cause intervalley scattering  but affects the spin of electrons. The Hamiltonian describing spin-orbit coupling at one of the Dirac points reads as \cite{Katsnelson}, \cite{Avsar}
\begin{equation}\label{Hsoc0}
H_{soc}=\left(\begin{array}{cccc}
\Delta&\pi^\dagger&0&0\\
\pi&-\Delta&-2i\lambda&0\\
0&2i\lambda&-\Delta&\pi^\dagger\\
0&0&\pi&\Delta
\end{array}\right).
\end{equation}
The basis of the bispinors is $(A^{\uparrow}_{K},B^{\uparrow}_{K},A^{\downarrow}_{K},B^{\downarrow}_{K})$ where the arrow denotes spin-up or spin-down component.
The diagonal term $\Delta$ corresponds to the intrinsic spin-orbital coupling. The breakdown of the mirror symmetry caused by perpendicular electric field, proximity of the substrate or by the curvature can induce the Rashba spin-orbit interaction $\lambda$, see \cite{Kane}, \cite{Zhe}, \cite{Huertas}, \cite{AndoSO}. The intrinsic coupling competes with the Rashba term in the opening of the gap in the spectrum. The gap induced by the nonvanishing $\Delta$ gets closed as long as $\lambda$ gets dominant \cite{Kane}.

The bilayer graphene consists of two layers of graphene that are close  to each other. The two layers can be mutually oriented in different ways. In Bernal stacking (also called $AB$ stacking), the layers are assembled in such a way that $A_2$ atoms of the upper layer sit just above the $B_1$ atoms of the lower layer. The atoms $A_1$ of the lower layer are then below the center of the hexagons of the upper lattice. The effective Hamiltonian for the low-energy particles can be written in the following form \cite{CastroNeto}, \cite{McCann}, \cite{Falco},
\begin{equation}\label{Hblg0}
H_{blg}=\left(\begin{array}{cccc}
-\Delta&\pi^\dagger&0&v_3\pi\\
\pi&-\Delta&v_1&0\\
0&v_1&\Delta&\pi^\dagger\\
v_3\pi^\dagger&0&\pi&\Delta
\end{array}\right).
\end{equation}
The basis of the bispinors is $(A_1,B_1,A_2,B_2),$ where $A_j (B_j)$ are $A(B)$ atoms from the $j$-th layer. The diagonal term $-\sigma_3\otimes\sigma_0\,\Delta$ reflects possible difference of the electrostatic potential in the two layers. The parameter $v_1$ is proportional to the $A_2-B_1$ hopping energy whereas $v_3$  is the coupling constant of the trigonal warping term. There holds $v_3\ll v_F$ and $v_3$ is frequently set to zero in the analysis of the low-energy regime \cite{Manes}.

The article is organized as follows. In the next section, we will discuss a class of coupled systems whose evolution equation can be reduced into two equations of lower dimension. First, we discuss the general framework and then we apply it on two-dimensional Dirac Hamiltonian. In the third section, we apply the scheme on the energy operator (\ref{Hdis0}) describing disorder scattering.  The Hamiltonians  (\ref{Hsoc0}) and (\ref{Hblg0}) are considered in the fourth section. Explicit examples are presented. The last section is left for discussion.

\section{Reducible coupled systems}

In order to introduce the general idea and the related notation, let us consider a generic quantum system which is described by the following evolution equation 
\begin{equation}\label{H}
\mathcal{H}\Psi=\left(-i\partial_t+\mathbf{1}\otimes D+\mathbb{S}_{11}\otimes V_{11}+\sum_{j,k=2}^N \mathbb{S}_{jk}\otimes V_{jk}\right)\Psi=0.
\end{equation}
The wave function $\Psi$ belongs to the space $\mathbb{C}^{N}\otimes \mathcal{L}$ where $\mathcal{L}$ is a Hilbert space whose actual form is not essential at the moment. The operator $\mathbf{1}$ is identity matrix on $\mathbb{C}^{N}$. The matrices $\mathbb{S}_{ij}$ form the basis of $N\times N$ matrices. They have all entries vanishing except a single matrix element,
\begin{equation}
(\mathbb{S}_{jk})_{lm}=\delta_{jl}\delta_{km},\quad j,k,l,m\in\{1,\dots,N\}.
\end{equation}
The operator $\mathbf{1}\otimes D$ represents a kinetic energy term. We can it leave unspecified at the moment. The potential term is required to be hermitian which imposes corresponding restriction on the operators $V_{jk}$ that are acting on $\mathcal{L}$. 

The potential term in (\ref{H}) is block-diagonal, i.e. the equation (\ref{H}) can be partially decoupled. It allows us to find some of its solutions by solving the reduced equation
\begin{equation}\label{Hred}
\left(-i\partial_t+D+V_{11}\right)\psi=0,
\end{equation}
where $\psi\in\mathcal{L}$. Indeed, if $\psi$ satisfies the equation above, then the wave function $\Psi=(\psi,0,\dots,0)^T$ is solution of (\ref{H}). This partial solvability is preserved in a large family of equations $\tilde{\mathcal{H}}\tilde{\Psi}=0$ that can be obtained from (\ref{H}) by a similarity transformation $\mathcal{U}$, $\tilde{\mathcal{H}}=\mathcal{U}\mathcal{H}\mathcal{U}^{-1}$. If we require $\mathcal{U}$ to preserve the form of the kinetic energy term and to keep the operator $\tilde{\mathcal{H}}$ hermitian, we can define the transformation as $\mathcal{U}=\mathbb{U}\otimes 1$ with $\mathbb{U}$ being a constant unitary matrix. The operator $\tilde{\mathcal{H}}$ can be called reducible as some of its zero modes can be obtained by solution of the reduced equation (\ref{Hred}). We can denote the number of reduced equations associated with (\ref{H}) as the degree of the reducibility of (\ref{H}). For instance the potential term $V=\mathbb{S}_{11}\otimes V_{11}+\mathbb{S}_{22}\otimes V_{22}+\sum_{j,k=3}^N \mathbb{S}_{jk}\otimes V_{jk}$ would possess reducibility of at least second degree.

Let us turn our attention to the special case where (\ref{H}) corresponds to a two-dimensional Dirac equation
\begin{equation}\label{HD}
\mathcal{H}\Psi=\left[-i\partial_t+\sigma_0\otimes (-i\sigma_1\partial_x-i\sigma_2\partial_y)+\sum_{k=1}^2\mathbb{S}_{k}\otimes\left(\begin{array}{cc}a_k&b_k\\b_k^\dagger&d_k\end{array}\right)\right]\Psi=0.
\end{equation}
For convenience, we denoted the projectors to the upper- and lower-spinor space as $\mathbb{S}_k\equiv \mathbb{S}_{kk}$. 
The equation (\ref{HD}) can be reduced into these two equations
\begin{eqnarray}\label{reducedequationD}
&&\left[-i\partial_t+\left(\begin{array}{cc}0&-i\partial_x-\partial_y\\-i\partial_x+\partial_y&0\end{array}\right)+\left(\begin{array}{cc}a_1&b_1\\b_1^\dagger& d_1\end{array}\right)\right]\left(\begin{array}{c}\psi_1\\\psi_2\end{array}\right)=0,\nonumber\\
&& \left[-i\partial_t+\left(\begin{array}{cc}0&-i\partial_x-\partial_y\\-i\partial_x+\partial_y&0\end{array}\right)+\left(\begin{array}{cc}a_2&b_2\\b_2^\dagger& d_2\end{array}\right)\right]\left(\begin{array}{c}\xi_1\\\xi_2\end{array}\right)=0.
\end{eqnarray}
Therefore, any solution of (\ref{HD}) can be written as a linear combination of 
\begin{equation}\label{PsiXi}
\Psi=(\psi_1,\psi_2,0,0)^T,\quad \Xi=(0,0,\xi_1,\xi_2)^T.
\end{equation}

Let us generate the class of reducible systems associated with (\ref{HD}) by the following unitary transformation $\mathcal{U}$,
\begin{equation}
\mathcal{U}= \mathbb{U}\otimes 1.
\end{equation}
We can fix the unitary matrix $\mathbb{U}$ without loss of generality as
\begin{equation} \mathbb{U}=\left(\begin{array}{cc}\cos\tau&-e^{-i\phi}\sin\tau\\e^{i\phi}\sin\tau&\cos\tau\end{array}\right),\quad \phi,\tau\in\mathbb{R},\quad \mathbb{U}^\dagger=\mathbb{U}^{-1}.
\end{equation}
The transformed equation reads as
\begin{equation}\label{reducibleH0}
\tilde{\mathcal{H}}\tilde{\Psi}=\mathcal{U}\mathcal{H}\mathcal{U}^{-1}\tilde{\Psi}=
\left[-i\partial_t+\sigma_0\otimes (-i\sigma_1\partial_x-i\sigma_2\partial_y)+\tilde{\mathcal{V}}\right]\tilde{\Psi}=0,\quad \tilde{\Psi}=\mathcal{U}\Psi,
\end{equation}
where
\begin{eqnarray}
&&\tilde{\mathcal{V}}=\sum_{k=1}^2\mathbb{U}\mathbb{S}_{k}\mathbb{U}^{-1}\otimes\left(\begin{array}{cc}a_k&b_k\\b_k^\dagger&d_k\end{array}\right)=\\
&&\left(
\begin{array}{cccc}
\cos^2\tau a_1+\sin^2\tau a_2 & \cos^2\tau b_1+\sin^2\tau b_2 
& 
\frac{1}{2}e^{-i \phi}\,(a_1-a_2)\sin 2\tau & 
\frac{1}{2}e^{-i \phi}(b_1-b_2)\sin 2\tau \\
\cos^2\tau b_1^\dagger+\sin^2\tau b_2^\dagger & 
\cos^2\tau d_1+\sin^2\tau d_2& 
\frac{1}{2}\, e^{-i \phi}(b_1-b_2)^\dagger\sin 2\tau& 
\frac{1}{2}e^{-i \phi}(d_1-d_2)^\dagger\sin 2\tau \\
\frac{1}{2}\, e^{i \phi} (a_1-a_2)^\dagger\, \sin 2\tau  & 
\frac{1}{2}\,e^{i \phi} (b_1-b_2)\, \sin 2\tau & 
\sin^2\tau a_1 +\cos^2\tau a_2& 
b_2\cos^2\tau+b_1\sin^2\tau \\
\frac{1}{2}e^{i \phi}(b_1-b_2)^\dagger\sin 2\tau& 
\frac{1}{2}e^{i \phi}(d_1-d_2)\sin 2\tau & (b_2\cos^2\tau+b_1\sin^2\tau)^\dagger& 
\sin^2\tau d_1+\cos^2\tau d_2
\end{array}
\right).
\label{Vreducible0}
\end{eqnarray}
We would like to match $\tilde{\mathcal{H}}$ with the corresponding Dirac operators in (\ref{Hdis0}), (\ref{Hsoc0}) or  (\ref{Hblg0}). First, we can see that the kinetic terms of (\ref{Hdis0}) and (\ref{reducibleH0}) does not coincide. When comparing (\ref{reducibleH0}) with (\ref{Hdis0}) or (\ref{Hblg0}), we can see that the potential term in (\ref{Vreducible0}) does not allow to set the components $V_{14}$ and $V_{23}$ independently. We can partially fix these deficiencies by an additional unitary transformation $\mathcal{R}$,
\begin{equation}
\mathcal{R}=\left(
\begin{array}{cccc}
1 & 0 & 0 & 0 \\
0 & 1 & 0 & 0 \\
0 & 0 & 0 & \epsilon \\
0 & 0 & 1 & 0
\end{array}
\right),\quad \epsilon=\pm 1,\quad \mathcal{R}^\dagger\mathcal{R}=1.
\end{equation}
The parameter $\epsilon$ makes it possible to choose between two different kinetic terms that appear in (\ref{Hdis0}) or (\ref{Hsoc0}) and (\ref{Hblg0}). The transformed equation reads as
\begin{equation}\label{reducibleH}
\underline{\mathcal{H}}\,\underline{\Psi}=\mathcal{R}\tilde{\mathcal{H}}\mathcal{R}^{-1}\underline{\Psi}=\left[-i\partial_t+\left(
\begin{array}{cccc}
0 & {\pi}  & 0&0 \\
{\pi} ^\dagger&0& 0 & 0 \\
0 & 0 & 0 & \epsilon\,{\pi}^\dagger  \\
0 & 0 &\epsilon\, {\pi} & 0
\end{array}
\right)+\underline{\mathcal{V}}\right]\underline{\Psi}=0,\quad \underline{\Psi}=\mathcal{R}\tilde{\Psi},
\end{equation}
where  
\begin{eqnarray}
\underline{\mathcal{V}}&=&\mathcal{R}\tilde{\mathcal{V}}\mathcal{R}^{-1}\nonumber\\&=&\left(
\begin{array}{cccc}
\cos^2\tau a_1+\sin^2\tau a_2 & \cos^2\tau b_1+\sin^2\tau b_2 
& 
\frac{1}{2}\epsilon\, e^{-i \phi}\,(b_1-b_2)\sin 2\tau & 
\frac{1}{2}e^{-i \phi}(a_1-a_2)\sin 2\tau \\
\cos^2\tau b_1^\dagger+\sin^2\tau b_2^\dagger & 
\cos^2\tau d_1+\sin^2\tau d_2& 
\frac{1}{2}\epsilon\, e^{-i \phi}(d_1-d_2)\sin 2\tau& 
\frac{1}{2}e^{-i \phi}(b_1-b_2)^\dagger\sin 2\tau \\
\frac{1}{2}\epsilon\, e^{i \phi} (b_1-b_2)^\dagger\, \sin 2\tau  & 
\frac{1}{2}\epsilon\,e^{i \phi} (d_1-d_2)\, \sin 2\tau & 
\sin^2\tau d_1 +\cos^2\tau d_2& 
\epsilon\,(b_2\cos^2\tau+b_1\sin^2\tau)^\dagger \\
\frac{1}{2}e^{i \phi}(a_1-a_2)\sin 2\tau& 
\frac{1}{2}e^{i \phi}(b_1-b_2)\sin 2\tau & \epsilon\,(b_2\cos^2\tau+b_1\sin^2\tau)& 
\sin^2\tau a_1+\cos^2\tau a_2
\end{array}
\right).\nonumber\\
\label{Vreducible1}
\end{eqnarray}
Let us notice that the unitary transformation relating $\mathcal{H}$ and $\underline{\mathcal{H}}$ is explicitly 
\begin{equation}
\mathcal{R}\,\mathcal{U}=\left(\begin{array}{cccc}
\cos\tau&0&-e^{-i\phi}\sin\tau&0\\
0&\cos\tau&0&-e^{-i\phi}\sin\tau\\
0&e^{i\phi}\epsilon\sin\tau&0&\epsilon \cos\tau\\
e^{i\phi}\sin\tau&0&\cos\tau&0
\end{array}\right).\label{totalU}
\end{equation}

There are ten independent matrix elements in a hermitian $4\times4$ matrix. In (\ref{Vreducible1}), we are free to fix six of them whereas the remaining four are determined by this choice already. For instance, we can fix the potential term in the following form
\begin{equation}
\underline{\mathcal{V}}=\left(
\begin{array}{cccc}
V_{11} 
& 
V_{12} & 
e^{-i\phi}V_{13}&e^{-i\phi}V_{14} \\
V_{12}^\dagger & 
V_{22}& 
e^{-i\phi}V_{23}& \epsilon\,e^{-i\phi} V_{13}^\dagger \\
e^{i\phi}V_{13}^\dagger & 
e^{i\phi} V_{23} & 
V_{22}-2\epsilon\, V_{23}\cot 2\tau & 
\epsilon\, V_{12}^\dagger-2 V_{13}^\dagger\cot 2\tau \\
e^{i\phi}V_{14}& 
\epsilon\, e^{i\phi} V_{13} &\epsilon\,V_{12}-2 V_{13}\cot 2\tau& 
V_{11}-2V_{14}\cot 2\tau
\end{array}
\right).
\label{Vreducible2}
\end{equation}
Here the components $V_{14}=\frac{1}{2}(a_1-a_2)\sin2\tau$ and $V_{23}=\frac{1}{2}(d_1-d_2)\sin2\tau$ are hermitian.
The reduced equations corresponding to (\ref{Vreducible2}) can be written as

\begin{eqnarray}\label{reducedequationD2}
&&\left[-i\partial_t+\left(\begin{array}{cc}0&\pi\\ \pi^\dagger&0\end{array}\right)
+
\left(\begin{array}{cc}V_{11}&V_{12}\\V_{12}^\dagger& V_{22}\end{array}\right)
+\epsilon\tan\tau\left(\begin{array}{cc}\epsilon\, V_{14}&V_{13}\\V_{13}^\dagger& V_{23}\end{array}\right)
\right]
\left(\begin{array}{c}\psi_1\\\psi_2\end{array}\right)=0,\nonumber\\
&&\left[-i\partial_t+\left(\begin{array}{cc}0&\pi\\ \pi^\dagger&0\end{array}\right)
+
\left(\begin{array}{cc}V_{11}&V_{12}\\V_{12}^\dagger& V_{22}\end{array}\right)
-\epsilon\cot\tau\left(\begin{array}{cc}\epsilon\,V_{14}&V_{13}\\V_{13}^\dagger& V_{23}\end{array}\right)
\right]
\left(\begin{array}{c}\xi_1\\\xi_2\end{array}\right)=0.
\end{eqnarray}
The two bispinors $\Psi$ and $\Xi$ get transformed into the following form
\begin{equation}\label{UPsiXi}
\underline{\Psi}=\mathcal{R}\,\tilde{\Psi}=\left(\begin{array}{r}
\cos\tau\, \psi_1\\
\cos\tau\,\psi_2\\
\epsilon\, e^{i\phi}\sin\tau\,\psi_2\\
e^{i\phi}\sin\tau\,\psi_1
\end{array}\right),\quad 
\underline{\Xi}=e^{i\phi}\mathcal{R}\,\tilde{\Xi}=\left(\begin{array}{r}
-\sin\tau\, \xi_1\\
-\sin\tau\, \xi_2\\
\epsilon\,e^{i\phi}\cos\tau\,\xi_2\\
\,e^{i\phi}\cos\tau\,\xi_1
\end{array}\right).
\end{equation}
By construction, the system described by (\ref{reducibleH}) is unitary equivalent to (\ref{HD}) so that the densities of probability as well as the transition amplitudes are not affected by the unitary mapping. Nevertheless, the transformed potential term (\ref{Vreducible1}) can represent a different physical situation when compared to (\ref{HD}). We will illustrate this point in the next sections on the explicit examples.

It is worth mentioning that there is a class of interactions that have vanishing expectation value on the states $\underline{\Psi}$ and $\underline{\Xi}$. Indeed, if we define
\begin{equation}\label{perturbation}
\delta \mathcal{V}=\mathbb{S}_{12}\otimes\left(\begin{array}{cc}v_{1}&v_{2}\\v_{3}&v_{4}\end{array}\right)+
\mathbb{S}_{21}\otimes\left(\begin{array}{cc}v_{1}^\dagger&v_{3}^\dagger\\v_{2}^\dagger&v_{4}^\dagger\end{array}\right),
\end{equation}
then it is rather direct observation that there holds 
$\Psi^\dagger \delta \mathcal{V}\Psi=\Xi^\dagger \delta \mathcal{V}\Xi=0,$
where $\Psi$ and $\Xi$ are given in (\ref{PsiXi}). This property is not altered by the unitary transformation  (\ref{totalU}). Therefore, we can write 
\begin{eqnarray}
\underline{\Psi}^\dagger \underline{\delta \mathcal{V}}\,\underline{\Psi}=\underline{\Xi^\dagger}\, \underline{\delta \mathcal{V}}\,\underline{\Xi}=0,\quad \mbox{where}\quad
\underline{\delta\mathcal{V}}=\mathcal{R}\,\mathcal{U}\delta\mathcal{V}\,\mathcal{U}^{-1}\mathcal{R}^{-1}.
\label{perturbationtransformed}
\end{eqnarray}
We will not present here the explicit form of $\underline{\delta\mathcal{V}}$ as it can be obtained in straightforward manner with the use of (\ref{totalU}) and (\ref{perturbation}). The property (\ref{perturbationtransformed}) can be important when  $\underline{\delta\mathcal{V}}$ corresponds to a perturbation potential. If $\underline{\Psi}$ (or $\underline{\Xi}$) corresponds to a bound state of $\underline{\mathcal{H}}$ with energy $E$, then the energy level is rather robust with respect to the small perturbation $\underline{\delta\mathcal{V}}$ as the first order correction term for energy is vanishing, 
$$\delta E=\langle \underline{\Psi}, \underline{\delta \mathcal{V}}\,\underline{\Psi}\rangle=0.$$
We will discuss the explicit form of $\underline{\delta \mathcal{V}}$ in the examples illustrated in the next section.

Let us conclude this section from two possible points of view. First, when we have an equation (\ref{HD}) with the potential term (\ref{Vreducible2}), we can make an ansatz (\ref{UPsiXi}) for the bispinors and get the two equations (\ref{reducedequationD2}). The simple form of (\ref{Vreducible2}) can serve for a quick test of reducibility of the potential term \footnote{In line with our comment below (\ref{Vreducible1}), it is possible to fix another six independent components and get another four dependent components in (\ref{Vreducible1}).}. Alternatively, we can construct the reducible model with $4\times4$ matrix potential (\ref{Vreducible1}) starting with the equations (\ref{reducedequationD}). This approach is more suitable for construction of reducible system by employing the known equations (\ref{reducedequationD}).

\section{Disorder scattering Hamiltonian}
Let us identify (\ref{Vreducible1}) with the potential term in (\ref{Hdis0}). We fix $\epsilon=-1$. There are two different ways how we can identify the matrix elements in the potential term in
\begin{equation}\label{disorderH}
\left[-i\partial_t+\left(
\begin{array}{cccc}
0 & \pi  & 0&0 \\
\pi ^\dagger&0& 0 & 0 \\
0 & 0 & 0 & -\pi^\dagger  \\
0 & 0 &- \pi & 0
\end{array}
\right)+\left(
\begin{array}{cccc}
V_A & V  & W_A&W^+ \\
\overline{V}&V_B& W^+ & W_B \\
\overline{W}_A & \overline{W}^+ & V_A & V'  \\
\overline{W}^+ & \overline{W}_B &\overline{V}' & V_B
\end{array}
\right)\right]\Psi=0.
\end{equation}

First, we fix
\begin{equation}\label{way1}
a_1=d_2,\quad a_2=d_1.
\end{equation}
Then
\begin{eqnarray}
&&V_A=\cos^2\tau d_2+\sin^2\tau d_1,\quad V_B=\cos^2\tau d_1+\sin^2\tau d_2\nonumber\\
&&V=\cos^2\tau b_1+\sin^2\tau b_2,\quad V'=-(b_2\cos^2\tau+b_1\sin^2\tau)^\dagger
\\
&&W_A=-\frac{1}{2}e^{-i \phi}(b_1-b_2)\sin 2\tau,\quad W_B=-e^{-2i\phi}W_A^\dagger
\\
&&W^+=\frac{1}{2}e^{-i \phi}(d_2-d_1)\sin 2\tau.
\end{eqnarray}
The two associated reduced equations are
\begin{eqnarray}\label{reduced3aDis}
&&\left[
-i\partial_t+\left(\begin{array}{cc}0&\pi\\\pi^\dagger&0\end{array}\right)
+\left(\begin{array}{cc}
d_2&b_1\\
b_1^\dagger&d_1
\end{array}\right)
\right]
\left(\begin{array}{c}\psi_1\\\psi_2\end{array}\right)=0,\\
&&\left[-i\partial_t+
\left(\begin{array}{cc}
0& \pi\\
\pi^\dagger&0
\end{array}\right)
+\left(\begin{array}{cc}
d_1&b_2\\b_2^\dagger&d_2\end{array}\right)\right]\left(\begin{array}{c}\xi_1\\\xi_2\end{array}\right)=0.\label{reduced3bDis}
\end{eqnarray}

The other option is to fix
\begin{equation}\label{way2}
a_1=d_1+d_2-a_2,\quad \tau=\pi/4.
\end{equation}
Then we get
\begin{eqnarray}
&&V_A=V_B=\frac{1}{2} (d_1+d_2),\quad V=\frac{1}{2}(b_1+ b_2),\quad V'=-\frac{1}{2}(b_1+ b_2)^\dagger,
\\
&&W_A=\frac{1}{2}e^{-i \phi}(b_2-b_1),\quad W_B=\frac{1}{2}e^{-i \phi}(b_1-b_2)^\dagger,\quad W^+=\frac{1}{2}e^{-i \phi}(d_1+d_2-2a_2). 
\end{eqnarray}
The two associated reduced equations are
\begin{eqnarray}\label{reduced3cDis}
&&\left[
-i\partial_t+\left(\begin{array}{cc}0&\pi\\\pi^\dagger&0\end{array}\right)
+\left(\begin{array}{cc}
d_1+d_2-a_2&b_1\\
b_1^\dagger&d_1
\end{array}\right)
\right]
\left(\begin{array}{c}\psi_1\\\psi_2\end{array}\right)=0,\\
&&\left[-i\partial_t+
\left(\begin{array}{cc}
0& \pi\\
\pi^\dagger&0
\end{array}\right)
+\left(\begin{array}{cc}
a_2&b_2\\b_2^\dagger&d_2\end{array}\right)\right]\left(\begin{array}{c}\xi_1\\\xi_2\end{array}\right)=0.\label{reduced3dDis}
\end{eqnarray}
Below, we present two examples of the reducible Hamiltonian (\ref{Hdis0}) where each of them corresponds to one of the proposed substitutions (\ref{way1}) or (\ref{way2}).

\subsection{P\"oschl-Teller system}\label{ptellersec}

First, we fix the potential terms in the equations (\ref{reduced3aDis}) and (\ref{reduced3bDis}) in accordance with (\ref{way1}) in the following manner
\begin{eqnarray}\label{PT1}
&&\left[-i\partial_t+\left(\begin{array}{cc}0&-i\partial_x+\partial_y\\-i\partial_x-\partial_y&0\end{array}\right)+\left(\begin{array}{cc}0&-i2\delta_1\tanh(x/2\delta_1)\\ i2\delta_1 \tanh(x/2\delta_1)&0\end{array}\right)\right]\left(\begin{array}{c}\psi_1\\\psi_2\end{array}\right)=0,\\
&& \label{h12}
\left[-i\partial_t+
\left(\begin{array}{cc}0&-i\partial_x+\partial_y\\-i\partial_x-\partial_y&0\end{array}\right)
+\left(\begin{array}{cc}0& -2\delta_2\tanh\left(y/2\delta_2\right) \\- 2 \delta_2\tanh\left(y/2\delta_2\right) &0\end{array}\right)\right]\left(\begin{array}{c}\xi_1\\\xi_2\end{array}\right)=0.\label{PT2}
\end{eqnarray}
Here we suppose that $\delta_1>0$ and $\delta_2>0$. The equations (\ref{PT1}) and (\ref{PT2}) can be mapped one into the other one by  an unitary transformation accompanied by an exchange of the variables $x\leftrightarrow y$ and of the coupling parameters $\delta_1\leftrightarrow \delta_2$.  Each of the two equations corresponds to the Dirac equation with P\"oschl-Teller potential that was solved analytically in \cite{Milpas}. Let us notice that an asymmetric version of the P\"oschl-Teller was studied in \cite{VJIshkhanyan}.

\begin{figure}[h]
	\centering
		\centering
		\includegraphics[scale=.5]{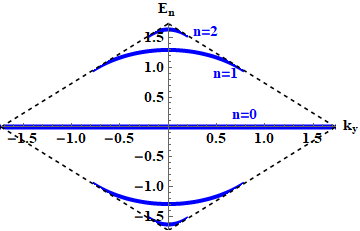}
	\caption{The energy bands $E_n=E_n(k_y)$ in (\ref{tellerspec}). The dashed lines are formed by the function $\pm |k_y-2\delta_1|$ and $\pm |k_y+2\delta_1|$. We fixed $\delta_1=\sqrt{3}/2$.
	}
	\label{tellerspecfig}
\end{figure}

The system described by (\ref{PT1}) has translational invariance, i.e. it is possible to fix the momentum $k_y$ along the symmetry axis and study the spectral properties of the effective one-dimensional system. There are discrete energies in the gap whose values are given by the following formula
\begin{equation}
E_n^{(\delta_1)}=\sqrt{2 \left(n-\frac{n^2}{8 \delta_1 ^2}\right) \left(1-\left(\frac{2\delta_1
		 k_y}{4\delta_1^2-n}\right)^2\right)}.
\label{tellerspec}
\end{equation}
The associated square integrable eigenvector can be written as 
\begin{equation}\label{tellermodes}
\psi_n=e^{i (-E_n t+k_y y)}\left(
\begin{array}{c}
-\frac{i}{E_n}(\partial_x+k_y+ 2\delta_1 z) \left(\frac{z+1}{2}\right)^{\sigma} \left(\frac{1-z}{2}\right)^{\rho} P_n^{2\rho,2 \sigma}(z) \\
\left(\frac{z+1}{2}\right)^{\sigma} \left(\frac{1-z}{2}\right)^{\rho} P_n^{2\rho,2 \sigma}(z)  \\
\end{array}
\right),
\end{equation}
where $\sigma=\delta_1\sqrt{(k_y-2 \delta_1 )^2-E_n^2}$, $\rho=\delta_1\sqrt{(2 \delta_1 +k_y)^2-E_n^2}$, $z=\tanh(x/2\delta_1)$. The requirement of square integrability of $\psi_n$ restricts both the range of positive integers $n$, $n\in\{0,\dots, 4\delta_1^2\}$ and the range of $k_y$, $|k_y|<2\delta_1$, see \cite{Milpas}. Additionally, the parameters $\sigma$ and $\rho$ have to be real. It implies that $E_n$ has to satisfy the following inequalities, $|k_y-2\delta_1|>|E_n|$, $|k_y+2\delta_1|>|E_n|$. These inequalities define the region where the discrete energy bands $E_n$ can appear, see Fig.~\ref{tellerspecfig}.
The energy spectrum of the Hamiltonian in (\ref{PT2}) and the corresponding solutions $(\xi_1,\xi_2)$ of (\ref{PT2}) can be obtained from (\ref{tellerspec}) and (\ref{tellermodes}) in the straightforward manner and we will not present them here explicitly.

\begin{figure}
	\centering 
	\begin{subfigure}[b]{0.35\textwidth}
		\centering
		\includegraphics[width=\textwidth]{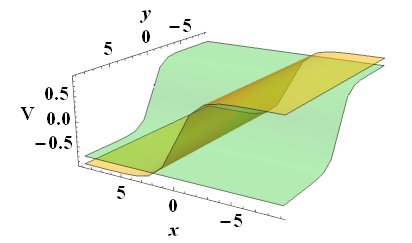}
	\end{subfigure}
\begin{subfigure}[b]{0.35\textwidth}
		\centering
		\includegraphics[width=\textwidth]{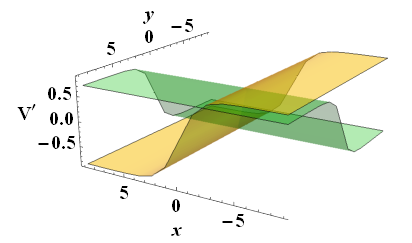}
		\label{fig2}
	\end{subfigure}
	\begin{subfigure}[b]{0.35\textwidth}
		\centering
		\includegraphics[width=\textwidth]{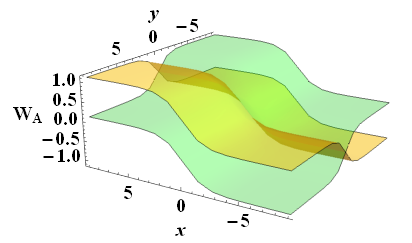}
		\label{fig2}
	\end{subfigure}
\begin{subfigure}[b]{0.35\textwidth}
		\centering
		\includegraphics[width=\textwidth]{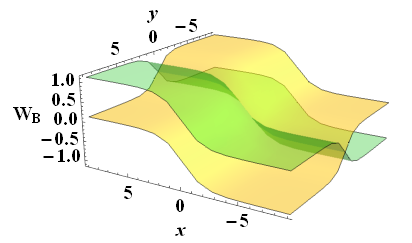}
		\label{fig2}
	\end{subfigure}
	\caption{This figure represents the real part (green) and the imaginary part (yellow) of the  expressions $V$, $V'$, $W_A$ and $W_B$ in the potential (\ref{disorderH}). Here we used the parameters $\delta_1=\sqrt{3}/2$, $\tau=\pi/4$, $\delta_2=1/\sqrt{2}$,  $\phi=\pi/4$}
	\label{Vcompfig}
\end{figure}

The reducible $4\times 4$ potential term (\ref{Vreducible1}) acquires the following form
\begin{equation}\label{disorderH}
\underline{\mathcal{V}}=\left(
\begin{array}{cccc}
 0 & V & W_A & 0 \\
 \overline{V}  & 0 & 0 & W_B \\
 \overline{W}_A & 0 & 0 & V' \\
 0 & \overline{W}_B & \overline{V}'& 0 \\
\end{array}
\right), 
\end{equation}

where 
\begin{eqnarray}
	V=-2i\delta_1 \cos^2\tau \tanh \frac{x}{2\delta_1}-2\delta_2\sin^2\tau\tanh\frac{y}{2\delta_2},&&\quad W_A=ie^{-i\phi}\sin 2\tau (\delta_1\tanh\frac{x}{2\delta_1}+i\delta_2\tanh \frac{y}{2\delta_2}),\\
	V'=-2i\delta_1 \sin^2\tau \tanh \frac{x}{2\delta_1}+2\delta_2\cos^2\tau\tanh\frac{y}{2\delta_2},&&\quad W_B=-e^{-2i\phi}W_A^\dagger.
\end{eqnarray}
The nonvanishing components of the potential term are illustrated in Fig.\ref{Vcompfig}.
%

\subsection{Crossed combs of scatterers }
\begin{figure}
	\centering
	\includegraphics[width=0.35\textwidth]{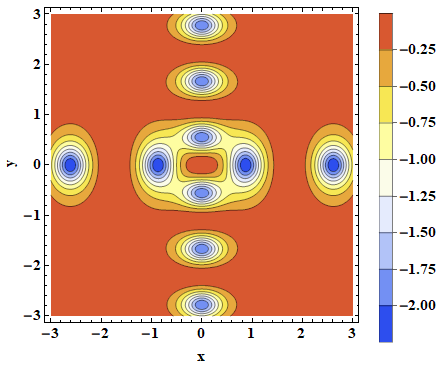} 	
	\includegraphics[width=0.35\textwidth]{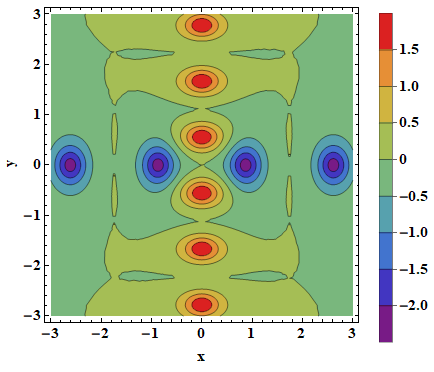}\\
	\includegraphics[width=0.35\textwidth]{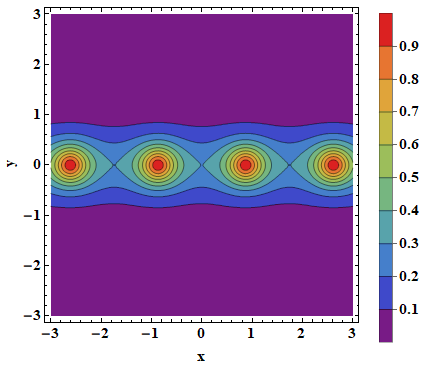}
	\includegraphics[width=0.35\textwidth]{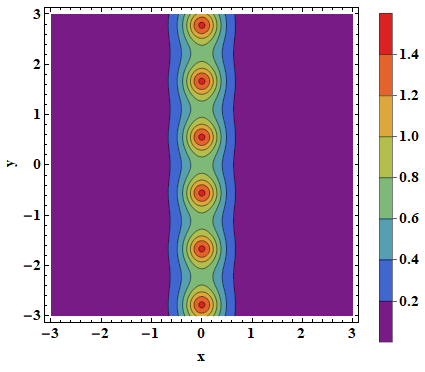}
	\caption{The figure illustrates $V_A$ (up left) and $W^+$ (up right) in (\ref{crosspotentials}), and density of probability of $\Psi$ (down left) and $\Xi$ (down right), see (\ref{UPsiXi}). In this case $m_2=2m_1=2$, $\omega_1=3/2$, $\omega_2=2$, $\phi=0$.}
	\label{fig3}
\end{figure}

The substitution (\ref{way2}) allows more freedom in the choice of the diagonal elements of the potential terms in (\ref{reduced3dDis}). For illustration, we pick up the model discussed recently in \cite{correa}. We fix the equations (\ref{reduced3cDis}) and (\ref{reduced3dDis}) as
\begin{eqnarray}
&&\left[-i\partial_t+\left(\begin{array}{cc}0&\pi \\\pi^\dagger&0\end{array}\right)-\left(\begin{array}{cc}\frac{4m_1\omega_1^2\sin^2\kappa_1x}{D_1(x,y)}&0\\0&\frac{4m_1\omega_1^2\sin^2\kappa_1x}{D_1(x,y)}\end{array}\right)\right]\left(\begin{array}{c}\psi_1\\\psi_2\end{array}\right)=0,\label{cc1}\\
&&\left[-i\partial_t+
\left(\begin{array}{cc}0& \pi \\ \pi^\dagger &0\end{array}\right)
-\left(\begin{array}{cc}\frac{4m_2\omega_2^2\sin^2\kappa_2y}{D_2(x,y)}&0\\0&\frac{4m_2\omega_2^2\sin^2\kappa_2y}{D_2(x,y)}\end{array}\right)\right]\left(\begin{array}{c}\xi_1\\\xi_2\end{array}\right)=0,\label{cc2}
\end{eqnarray}
where 
$$D_1(x,y)= m_1^2+\omega_1^2 \cos (2 \kappa_1  x)+\kappa_1 ^2 \cosh (2 \omega_1 y), \quad D_2(x,y)=m_2^2+\omega_2^2 \cos (2 \kappa_2  y)+\kappa_2 ^2 \cosh (2 \omega_2 x),$$ 
$\omega_a,$ $m_a$ are real constants and $\kappa_a=\sqrt{m_a^2+\omega_a^2}$, $a=1,2$. Let us notice that the equation (\ref{cc1}) can be transformed into (\ref{cc2}) by a matrix unitary transformation $\exp(i\pi\sigma_3/4)$ accompanied by the change of coordinates $x\rightarrow -y$, $y\rightarrow x$ and possibly different choice of the free parameters  $\omega_1\rightarrow\omega_2$, $\kappa_1\rightarrow\kappa_2$. The system described by (\ref{cc1}) can be solved for $E=m$. For this energy level, it is possible to find scattering states as well as two states localized at the potential barrier \cite{correa}. As the two localized states are practically identical when comparing their densities of probabilities, let us present just one of them. Its explicit form is
\begin{equation}\label{psicross}
\psi=\frac{\sqrt{2} \kappa ^2}{D_1(x,y)}e^{im_1t}
\left(\begin{matrix} \omega_1  \sin(\kappa_1  x) \sinh (\omega_1  y)-i \cosh (\omega_1  y) (m_1 \sin (\kappa_1  x)+\kappa_1  \cos (\kappa_1  x))\\
\cosh (\omega_1  y) (m_1 \sin (\kappa_1  x)-\kappa_1  \cos (\kappa_1  x))+i \omega_1  \sin (\kappa_1  x) \sinh (\omega_1  y) 
\end{matrix}\right),
\end{equation}
and it solves (\ref{cc1}). The corresponding solution of (\ref{cc2}) can be found in the following form
\begin{equation}\label{xicross}
\xi=e^{i\frac{\pi}{4}\sigma_3}\psi_1|_{x\rightarrow -y, y\rightarrow x, \omega_1\rightarrow\omega_2,\kappa_1\rightarrow\kappa_2}.
\end{equation}
The reducible potential $\underline{\mathcal{V}}$ in (\ref{Vreducible1}) then reads as
\begin{equation}\label{Vcross}
\underline{\mathcal{V}}=\left(
\begin{array}{cccc}V_A&0&0&W^+\\0&V_A&-W^+&0\\
0&-\overline{W^+}&V_A&0\\
\overline{W^+}&0&0&V_A
\end{array}
\right),
\end{equation}
where
\begin{equation}\label{crosspotentials}
V_A=-\frac{2m_1\omega_1\sin^2\kappa_1x}{D_1(x,y)}-\frac{2m_2\omega_2\sin^2\kappa_2x}{D_2(x,y)},\quad W^+=-2e^{-i\phi} \left(\frac{2m_1\omega_1\sin^2\kappa_1x}{D_1(x,y)}-\frac{2m_2\omega_2\sin^2\kappa_2x}{D_2(x,y)}\right).
\end{equation}
The localized solutions for the reducible system with the potential (\ref{Vcross}) can be constructed by substitution of (\ref{psicross}) and (\ref{xicross}) into (\ref{UPsiXi}). Their density of probability of the localized states  as well as the plots of the nonvanishing components of $\underline{\mathcal{V}}$ can be found in Fig.~\ref{fig3}. 

\section{Spin-orbit interaction and bi-layer graphene}
We fix $\epsilon=1$ and
\begin{equation}
a_2=a_1,\quad b_2=b_1=0,\quad \tau=\pi/4.
\end{equation}
Substituting these quantities into (\ref{Vreducible1}), we get the following equation for the reducible system,
\begin{equation}
\left[-i\partial_t+\left(
\begin{array}{cccc}
0 & \pi  & 0&0 \\
\pi ^\dagger&0& 0 & 0 \\
0 & 0 & 0 & \pi^\dagger  \\
0 & 0 &\pi & 0
\end{array}
\right)+\left(
\begin{array}{cccc}
\mu+\Delta & 0 & 0 & 0 \\
0 & \mu-\Delta &e^{-i\phi}\lambda & 0 \\
0 & e^{i\phi}\lambda& \mu-\Delta & 0 \\
0 & 0 & 0 & \mu+\Delta
\end{array}
\right)\right]\Psi=0,\label{Hso}
\end{equation}
where
\begin{equation}
\Delta=\frac{a_1}{2}-\frac{d_1+d_2}{4},\quad \mu=\frac{a_1}{2}+\frac{d_1+d_2}{4},\quad \lambda=\frac{d_1-d_2}{2}.
\end{equation}
In order to identify the kinetic energy term with (\ref{Hsoc0}) or (\ref{Hblg0}), we have to set
\begin{equation} \pi=-i\partial_x.\label{longitudinal0}\end{equation} 
This way, we are restricted to the effectively $1+1$ dimensional Hamiltonian. Such operator can be obtained by fixing $k_y=0$ in the two dimensional systems with translational symmetry along $y$ axis. We will suppose that this is the case. For $\phi=\pi/2$, the potential term in (\ref{Hso}) can be identified with the Hamiltonian with spin-orbital interaction (\ref{Hsoc0}). When $\phi=0$, it corresponds to the bilayer Hamiltonian (\ref{Hblg0}) with an additional inhomogeneous mass term on the two layers. 
The two associated reduced equations are
\begin{eqnarray}
\label{crossed1}
&&\left[
-i\partial_t+\left(\begin{array}{cc}0& \pi\\\pi^\dagger&0\end{array}\right)
+\left(\begin{array}{cc}
a_1&0\\
0&d_1
\end{array}\right)
\right]
\left(\begin{array}{c}\psi_1\\\psi_2\end{array}\right)=0,\\
&&\left[-i\partial_t+
\left(\begin{array}{cc}
0& \pi\\
\pi^\dagger&0
\end{array}\right)
+\left(\begin{array}{cc}
a_1&0\\0&d_2\end{array}\right)\right]\left(\begin{array}{c}\xi_1\\\xi_2\end{array}\right)=0.\label{reduced3bSoc}
\end{eqnarray}

\begin{figure}
	\centering
	\begin{subfigure}[b]{0.32\textwidth}
		\centering
		\includegraphics[width=\textwidth]{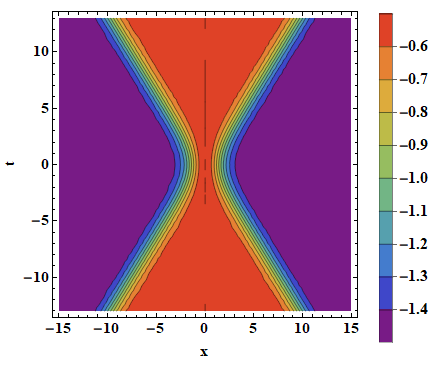}
	\end{subfigure} 
	\begin{subfigure}[b]{0.32\textwidth}
		\centering
		\includegraphics[width=\textwidth]{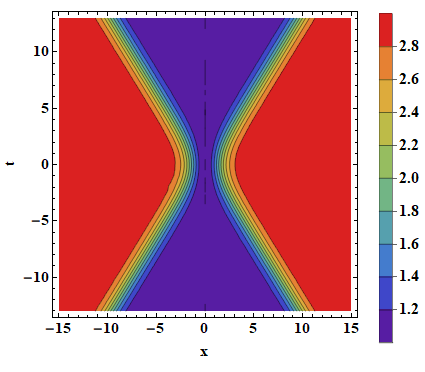}
	\end{subfigure} 
	\begin{subfigure}[b]{0.32\textwidth}
		\centering
		\includegraphics[width=\textwidth]{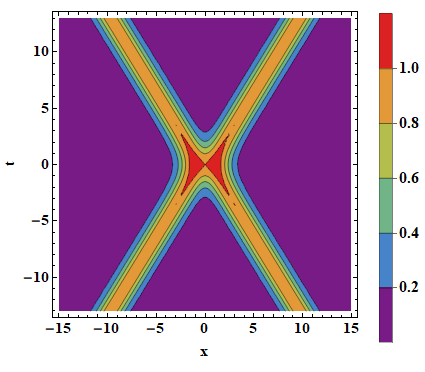}
	\end{subfigure}
	\caption{Plots of $\mu$ (left), $\lambda$ (center) from (\ref{mulambda}), and  the probability density (right) of $\Psi_1$ in (\ref{Psi}). We fixed $\Delta=1$, $m=1/2$, $\omega=1/2$, $\tau=\pi/4$.}
	\label{fig_time}
\end{figure}

When only one of the equations in (\ref{reduced3bSoc}) is exactly solvable, the system described in (\ref{Hso}) is quasi-exactly solvable as we can find only a part of the existing solutions analytically. In order to illustrate compatibility of the scheme with time-dependent potentials, we fix
\begin{equation}
a_1=m \left(-1+\frac{4 \kappa^2 \cosh ^2\omega t}{D(t,x)}\right),\quad d_1=-a_1,
\end{equation}\begin{equation}
D(t,x)=m^2+\kappa^2 \cosh (2 \omega t)+\omega^2 \cosh (2\kappa x),\quad \kappa=\sqrt{m^2+\omega^2}.
\end{equation}
We can leave $d_2$ unspecified as we are interested in the solutions of (\ref{crossed1}). With this choice of the potential functions, (\ref{crossed1}) is exactly solvable, see \cite{Samsonov}, \cite{correa}. Therefore, we can find solutions of the equation (\ref{Hso})  
where
\begin{equation}
\Delta=\frac{3a_1-d_2}{4},\quad \mu=\frac{a_1+d_2}{4} ,\quad 
\lambda= -\frac{a_1+d_2}{2}. 
\end{equation}
We can use the freedom in selecting the function $d_2=d_2(x,t)$ to fine-tune  the electrostatic field $\mu$, the intrinsic  spin-orbit interaction $\Delta$ and Rashba term $\lambda$. For instance, when  \begin{equation}d_2=-\left(4\delta+3m-\frac{12m\kappa^2 \cosh^2\omega t}{D(t,x)}\right),\end{equation} 
then $\Delta\equiv\delta$ is constant whereas $\mu$ and $\lambda$ are inhomogeneous both in space and time,
\begin{equation}\label{mulambda}
\mu=-\Delta-m+\frac{4m \kappa^2\cosh^2t\omega}{D(t,x)},\quad \lambda=2\left(\Delta+m-\frac{4m \kappa^2\cosh^2t\omega}{D(t,x)}\right),
\end{equation}
see Fig.~\ref{fig_time} for illustration.  Independently on the explicit choice of $d_2$, we can find a set of solutions that stem from the equation (\ref{crossed1}). 
For illustration, we present explicit form of two such states that solve the equation (\ref{Hso}),
\begin{equation}\label{Psi}
\Psi_1=\frac{1}{D(t,x)}\left(\begin{array}{l}
\sinh x \kappa (m\cosh t\omega+i\omega \sinh t\omega)\\
-i\kappa \cosh t\omega\cosh\kappa x\\
-ie^{i\phi}\kappa\cosh t\omega\cosh x\kappa\\
e^{i\phi}(m\cosh t\omega+i\omega \sinh t\omega)\sinh\kappa x
\end{array}\right),
\quad 
\Psi_2=\frac{1}{D(t,x)}\left(\begin{array}{l}
\kappa  \cosh t\omega \cosh \kappa x\\
(-i m \cosh t\omega-\omega \sinh t\omega)\sinh\kappa x\\
e^{i\phi}(-i m \cosh t\omega-\omega \sinh t\omega)\sinh\kappa x\\
e^{i\phi}\kappa \cosh t\omega \cosh \kappa x
\end{array}\right).
\end{equation}
The states are spatially strongly localized. We present their density of probability as a function of time in Fig.\ref{fig_time}.

In another scenario, we fix the functions $a_1$, $d_1$ and $d_2$ in the following manner,
\begin{equation}\label{scenarioII}
a_1=m_1(x)+V_1=m_2(x)+V_2,\quad d_1=-m_1(x)+V_1,\quad d_2=-m_2(x)+V_2=-m_1(x)-V_1+2V_2.
\end{equation}
The first equation implies that $m_2(x)=m_1(x)+V_1-V_2$.
Then the components of the potential term in (\ref{Hso}) acquire the following form,
\begin{equation}
\Delta(x,t)=m_1(x,t)+\frac{V_1-V_2}{2},\quad \mu=\frac{V_1+V_2}{2},\quad \lambda= V_1-V_2.
\end{equation}
The model presented in the Section~\ref{ptellersec} can serve us for a quick illustration. The equations (\ref{PT1}) and (\ref{PT2}) can be brought into the following form by a simple unitary transformation,
\begin{eqnarray}\label{PT3}
&&\left[-i\partial_t+\left(\begin{array}{cc}0&-i\partial_x\\-i\partial_x&0\end{array}\right)+\left(\begin{array}{cc}2\delta\tanh\frac{x}{2\delta}+k_y&0\\0&-2\delta\tanh\frac{x}{2\delta}-k_y\end{array}\right)\right]\psi=0,\\
&&\left[-i\partial_t+\left(\begin{array}{cc}0&-i\partial_x\\-i\partial_x&0\end{array}\right)+\left(\begin{array}{cc}2\delta\tanh\frac{x}{2\delta}+k_y-V_2&0\\ 0&-2\delta\tanh\frac{x}{2\delta}-k_y+V_2\end{array}\right)+V_2\right]\xi=0.\label{PT4}
\end{eqnarray}
Notice that $k_y$ does not play the role of longitudinal momentum as it is fixed to be vanishing, see (\ref{longitudinal0}). Here, $k_y$ corresponds just to an interaction parameter.  
When comparing with (\ref{scenarioII}), we identified $m_1=2\delta\tanh(x/2\delta)+k_y$ and $V_1=0$. The solutions of (\ref{PT3}) can be obtained easily from those discussed in the Section~ \ref{ptellersec}. We get
	\begin{equation}\label{tellermodesspin}
	{\psi}_n=e^{-i E_n t}U^{-1}\left(
	\begin{array}{c}
	-\frac{i}{E_n}(\partial_x+k_y+ 2\delta z) \left(\frac{z+1}{2}\right)^{\sigma} \left(\frac{1-z}{2}\right)^{\rho} P_n^{2\rho,2 \sigma}(z) \\
	\left(\frac{z+1}{2}\right)^{\sigma} \left(\frac{1-z}{2}\right)^{\rho} P_n^{2\rho,2 \sigma}(z) 
	\end{array}
	\right),\quad U=e^{i\frac{\pi}{4}\sigma_1}.
	\end{equation}
The parameters are given by $\sigma=\delta\sqrt{(k_y-2 \delta )^2-E_n^2}$, $\rho=\delta\sqrt{(k_y+2 \delta)^2-E_n^2}$, $z=\tanh(x/2\delta)$. The corresponding energies are 
\begin{equation}
E_n=\sqrt{2 \left(n-\frac{n^2}{8 \delta ^2}\right) \left(1-\left(\frac{2\delta k_y}{4\delta^2-n}\right)^2\right)}.
\label{tellerspecspin}
\end{equation}
The solution of (\ref{PT4}) can be obtained via a change of parameters and an additional phase factor, 
\begin{equation}
{\xi}_n=e^{-iV_2t}{\psi}_n|_{k_y\rightarrow k_y-V_2}.
\end{equation}
The reducible system with the potential in (\ref{Hso}) has the following explicit components, 
\begin{equation}
\Delta=k_y-\frac{V_2}{2}+2 \delta \tanh \frac{x}{2 \delta},\quad \mu= \frac{V_2}{2},\quad \lambda=-V_2.
\end{equation}
Therefore, there is a constant electrostatic field, constant Rashba interaction and space-dependent intrinsic spin-orbit term in the system. The later term acquires constant values asymptotically.

At the end of the section, let us notice that for the spin-orbit interaction, the potential term $\underline{\delta \mathcal{V}}$ in (\ref{perturbationtransformed}) reads as
	\begin{equation}
	\underline{\delta\mathcal{V}}|_{\epsilon=1,\tau=\frac{\pi}{4},\phi=\frac{\pi}{2}}=\left(\begin{array}{cccc}
	\mbox{Im}\, v_1 &\frac{i}{2}(\overline{v_3}-v_2)&\frac{1}{2}(v_2+\overline{v_3}) &\mbox{Re}\, v_1 \\
	-\frac{i}{2}(v_3-\overline{v_2})&\mbox{Im}\, v_4& \mbox{Re}\, v_4& \frac{1}{2}(\overline{v}_2+v_3)\\
	\frac{1}{2}(\overline{v_2}+v_3)&\mbox{Re}\,v_4&-\mbox{Im}\,v_4&-\frac{i}{2}(\overline{v_2}-v_3)\\
	\mbox{Re}\,v_1&\frac{1}{2}(v_2+\overline{v_3})&\frac{i}{2}(v_2-\overline{v_3})&-\mbox{Im}\, v_1
	\end{array}\right).\label{perturbationSO}
	\end{equation}
In case of bilayer graphene, we have
	\begin{equation}
	\underline{\delta\mathcal{V}}|_{\epsilon=1,\tau=\frac{\pi}{4},\phi=0}=\left(\begin{array}{cccc}
	-\mbox{Re}\, v_1 &-\frac{1}{2}(\overline{v_3}+v_2)&\frac{1}{2}(v_2-\overline{v_3}) &i\,\mbox{Im}\, v_1 \\
	-\frac{1}{2}(v_3+\overline{v_2})&-\mbox{Re}\, v_4& i\, \mbox{Im}\, v_4& \frac{1}{2}(-\overline{v}_2+v_3)\\
	\frac{1}{2}(\overline{v_2}-v_3)&-i\,\mbox{Im}\,v_4&\mbox{Re}\,v_4&\frac{1}{2}(\overline{v_2}+v_3)\\
	-i\,\mbox{Im}\,v_1&\frac{1}{2}(-v_2+\overline{v_3})&\frac{1}{2}(v_2+\overline{v_3})&\mbox{Re}\, v_1
	\end{array}\right).\label{perturbationblg}
	\end{equation}
In particular, fixing $v_1=v_4=0$, $v_3=-\overline{v_2}$, $v_2(x)\in\mathbb{R}$, the expression (\ref{perturbationSO}) reduces into
	\begin{equation}
		\underline{\delta\mathcal{V}}|_{\epsilon=1,\tau=\frac{\pi}{4},\phi=\frac{\pi}{2}}=v_2(x)\,\sigma_0\otimes\sigma_2.
	\end{equation}
It corresponds to the vector potential of the magnetic field perpendicular to the surface. Therefore, we can conclude that the discrete energies of the Hamiltonian in (\ref{Hso}) are robust with respect to small perturbations of the potential term by external magnetic field, or, identifying $v_2(x)\equiv \kappa_y$, with respect to small fluctuations of the longitudinal momentum $\kappa_y$.

\section{Conclusion}
In the article, we focused on the analysis of coupled quantum systems where exact solution of their dynamical equation can be obtained from lower-dimensional, uncoupled settings. In the presented reduction scheme, the family of reducible Dirac Hamiltonians was found in (\ref{reducibleH})-(\ref{Vreducible1}). We showed that its members are, by construction, related to an uncoupled system via unitary transformation.   
We compared the class of reducible Hamiltonians with the operators that describe physically relevant situations in graphene. In particular, we discussed the system with disorder scattering Hamiltonian in section 3 where the impurities or interaction with the substrate can cause intervalley scattering. The system with spin-orbit interaction or the model of bilayer graphene were considered in the section 4. In all these cases, we showed on explicit examples that the reduction scheme is applicable for potentials that lack translational symmetry as well as for the systems that depend explicitly on time.

The two Dirac operators, despite being related by unitary transformation, can describe very different physical systems. Contrary to the initial uncoupled operator, the unitary transformed Hamiltonian can serve in description of Dirac systems where either valley or spin-degrees of freedom are coupled by corresponding interactions. It is worth mentioning in this context that unitary transformation was used in \cite{Jakubsky1} to explain absence of back-scattering of Dirac fermions on electrostatic barriers. It was also employed in \cite{Jakubsky2} where exactly solvable model of Dirac fermions in electro-magnetic field was constructed.

We believe that the reducibility scheme can be useful for the analysis of the physics of graphene with the use of the exactly solvable models. In this context, let us mention e.g. analysis of the Dirac fermions in graphene in presence of an inhomogenous Rashba term \cite{Alomar}, \cite{Razzaghi}, \cite{Hosseni}, of $p-n$ or $p-n-p$ junctions with spin-orbit interaction \cite{Rataj}, \cite{Peeters}, or the recent analysis of the exactly solvable models of bilayer graphene \cite{FernandezBLG1}, \cite{FernandezBLG2}.  It would be also interesting to analyze $4\times 4$ systems that are not covered by the reducibility scheme. In this context, supersymmetric techniques could be particularly useful \cite{Schulze}. Finally, let us mention that in the current work, we supposed that the Dirac fermions live on entire plane. When terminated lattices are considered, the wave functions describing Dirac fermions are subject to specific boundary conditions. It would be interesting to discuss compatibility of the class of reducible Dirac operators with the boundary conditions. Nevertheless, we find further analysis in these directions beyond the scope of the present article.

\section*{Acknowledgement} 
M.C.-C. thanks Department of Physics of the Nuclear Physics Institute of CAS for hospitality. V. J. was supported by GA\v CR grant no 19-07117S. M.C.-C. acknowledges the support of CONACYT, project FORDECYT-PRONACES/61533/2020. M.C.-C. also acknowledges the Conacyt fellowship 301117.


\begin{thebibliography}{99}
	
\bibitem{Semenoff} W. G. Semenoff, "Condensed-Matter Simulation of a Three-Dimensional Anomaly," Phys. Rev. Lett {\bf53}, 2449 (1984).


\bibitem{Altland}A. Altland, "Low-energy theory of disordered graphene," Phys. Rev. Lett. {\bf 97}, 236802 (2006). 

\bibitem{Kechedzhi} E. McCann, K. Kechedzhi, V. I. Fal'Ko, H. Suzuura, T. Ando, B. L. Altshuler,  "Weak-Localization Magnetoresistance and Valley Symmetry in Graphene," Phys. Rev. Lett. {\bf97}, 146805 (2006).

\bibitem{Manes} J. L. Ma\~nes, F. Guinea, M. A. H. Vozmediano, "Existence and topological stability of Fermi points in multilayered graphene," Phys. Rev. B {\bf75}, 155424 (2007).

\bibitem{Shon}N. Shon, T. Ando, "Quantum Transport in Two-Dimensional Graphite System,"
J. Phys. Soc. Jap. {\bf67}, 2421 (1998).

\bibitem{AndoDisorderScatt} T. Ando, T. Nakanishi, "Impurity Scattering in Carbon Nanotubes: Absence of Back Scattering," J. Phys. Soc. Jap. {\bf67}, 1704 (1998). 

\bibitem{Frassdorf}Ch. Fr\"a\ss dorf, L. Trifunovic, N. Bogdanoff, P. W. Brouwer, "Graphene p n junction in a quantizing magnetic field: Conductance at intermediate disorder strength," Phys. Rev. B {\bf94}, 195439 (2016).

\bibitem{AndoCross}Ando, Tsuneya, "Crossover between Positive and Negative Magnetoresistance in Graphene: Roles of Absence of Backscattering," J. Phys. Soc. Jap. {\bf 90},  044712 (2021).



\bibitem{Katsnelson} M. I. Katsnelson, \textit{Graphene: Carbon in two dimensions}, Cambridge University Press 2012. 

\bibitem{Avsar} A. Avsar, H. Ochoa, F. Guinea, B. \"Ozyilmaz, B. J. van Wees, I. J. Vera-Marun, 
"Colloquium: Spintronics in graphene and other two-dimensional materials," Rev. Mod. Phys. {\bf92}, 021003  (2020).

\bibitem{Kane}C. L. Kane, E. J.Mele, "Quantum Spin Hall Effect in Graphene," Phys. Rev. Lett. {\bf95}, 226801,  (2005).

\bibitem{Zhe}Zhe Wang et al. 
"Strong interface-induced spin-orbit interaction in graphene on WS2,"	Nat. Commun.  {\bf6}, 8339 (2015 Sep 22).  

\bibitem{Huertas}D. Huertas-Hernando, F. Guinea, A. Brataas, "Spin-orbit coupling in curved graphene, fullerenes, nanotubes, and nanotube caps," 	Phys. Rev. B  {\bf74}, 155426 (2006).

\bibitem{AndoSO} T. Ando, "Spin-Orbit Interaction in Carbon Nanotubes," J. Phys. Soc. Jap. {\bf69}, 1757 (2000).


\bibitem{CastroNeto}A. H. Castro Neto, F. Guinea, N. M. R. Peres, K. S. Novoselov, A. K. Geim, "The electronic properties of graphene,"	Rev. Mod. Phys. {\bf81}, 109 (2009)

\bibitem{McCann}E. McCann, M. Koshino, "The electronic properties of bilayer graphene," Rep. Prog. Phys. {\bf 76},  056503 (2013).

\bibitem{Falco}E. McCann, D. S. L. Abergel, V. I. Fal'ko, "Electrons in bilayer graphene," Solid State Communications {\bf143}, 110 (2007).

\bibitem{Milpas}E. Milpas, M. Torres, G. Murgu\'i{}a, "Magnetic field barriers in graphene: an analytically solvable model," J. Phys. Condens. Matter {\bf23}, 245304 (2011).

\bibitem{VJIshkhanyan}A. M. Ishkhanyan, V. Jakubsk\'y, "Two-dimensional Dirac fermion in presence of an asymmetric vector potential," J. Phys A: Math. Theor. {\bf51}, 495205 (2018).

 \bibitem{correa} A. Contreras-Astorga, F. Correa, V. Jakubsk\'y, "Super-Klein tunneling of Dirac fermions through electrostatic gratings in graphene," Phys. Rev. B {\bf102},  115429 (2020).
 
 \bibitem{Samsonov}A. A. Pecheritsyn, E. O. Pozdeeva, B. F. Samsonov, "Darboux Transformation of the Nonstationary Dirac Equation,"  Russian Physics Journal {\bf48},  365 (2005).
 
 
 \bibitem{Jakubsky1}V. Jakubsk\'y, L.-M. Nieto, M. S. Plyushchay, Klein tunneling in carbon nanostructures: A free-particle dynamics in disguise, Phys. Rev. D  {\bf83}, 047702 (2011).
 
 \bibitem{Jakubsky2}V. Jakubsk\'y, "Spectrally isomorphic Dirac systems: Graphene in an electromagnetic field,"  
 Phys. Rev. D {\bf91}, 045039 (2015).
 

	
\bibitem{Alomar}M. I. Alomar, D. S\'anchez, "Thermoelectric effects in graphene with local spin-orbit interaction," Phys. Rev. B {\bf89}, 115422 (2014); erratum Phys. Rev. B {\bf91}, 039905 (2015).

\bibitem{Razzaghi}Razzaghi, Leila; Hosseini, Mir Vahid, "Quantum transport of Dirac fermions in graphene with a spatially varying Rashba spin-orbit coupling," Physica E: Low-dimensional Systems and Nanostructures {\bf 72}, 89 (2015).

\bibitem{Hosseni}Mir Vahid Hosseini, "The influence of anisotropic Rashba spin-orbit coupling on current-induced spin polarization in graphene," J. Phys. Condens Matter {\bf29}, 315502 (2017).

\bibitem{Rataj}M. Rataj, J. Barna\'s, "Graphene p-n junctions with nonuniform Rashba spin-orbit coupling," Appl. Phys. Lett. {\bf99}, 162107 (2011).

\bibitem{Peeters}Kh. Shakouri, M. Ramezani Masir, A. Jellal, E. B. Choubabi, and F. M. Peeters, "Effect of spin-orbit couplings in graphene with and without potential modulation," Phys. Rev. B {\bf88}, 115408 (2013).

\bibitem{FernandezBLG1}D. J. Fern\'andez, J. D. Garc\'i{}a, D. O-Campa, "Electron in bilayer graphene with magnetic fields leading to shape invariant potentials," J. Phys. A: Math. Theor. {\bf53}, 435202 (2020).

\bibitem{FernandezBLG2}David J. Fern\'andez, Juan D. Garc\'i{}a, D. O-Campa,"  
Bilayer graphene in magnetic fields generated by supersymmetry," arXiv:2101.05391

\bibitem{Schulze}E. Pozdeeva, A. Schulze-Halberg, "Darboux transformations for a generalized Dirac equation in two dimensions," J. Math. Phys. {\bf51}, 113501 (2010).
	
\end{thebibliography}
\end{document}